\documentclass[aps,twocolumn,superscriptaddress,amsfont,graphicx,nofootinbib,preprintnumbers]{revtex4-2}%
\usepackage{graphicx}
\usepackage{dcolumn}
\usepackage{bm}

\usepackage{subfigure}
\usepackage{color,epsfig}
\usepackage{ifpdf}
\usepackage{amsmath}
\usepackage[english]{babel}
\usepackage{amssymb}
\usepackage{braket}
\usepackage{hyperref}
\usepackage{enumerate}

\def\mev{~{\rm MeV}}

\def\gsim{\raise0.3ex\hbox{$>$\kern-0.75em\raise-1.1ex\hbox{$\sim$}}}
\def\lsim{\raise0.3ex\hbox{$<$\kern-0.75em\raise-1.1ex\hbox{$\sim$}}}
\usepackage{url}

\usepackage{slashed}

\usepackage{changes}
\begin{document}


\title{Spin-dependent Scattering of Boosted Dark Matter}

\author{Wenyu Wang}
\email{wywang@bjut.edu.cn}
\affiliation{Faculty of Science, Beijing University of Technology}

\author{Lei Wu}
\email{leiwu@njnu.edu.cn}
\affiliation{Department of Physics and Institute of Theoretical Physics, Nanjing Normal University, Nanjing, 210023, China}

\author{Wen-Na Yang}
\email{yangwenna22@mails.ucas.ac.cn
}
\affiliation{Faculty of Science, Beijing University of Technology}
\affiliation{School of Fundamental Physics and Mathematical Sciences, Hangzhou Institute for Advanced Study,~UCAS, Hangzhou 310024, China\\
University of Chinese Academy of Sciences, 100190 Beijing, China}

\author{Bin Zhu}
\email{zhubin@mail.nankai.edu.cn}
\affiliation{School of Physics, Yantai University, Yantai 264005, China}

\date{\today}

\begin{abstract}
Boosted dark matter is a promising method for probing light dark matter,  with a well-developed computational framework for spin-independent scattering already existing. 
The spin-dependent case, on the other hand, lacks a coherent treatment. 
We therefore give the first comprehensive derivation of the spin-dependent scattering cross-section for boosted dark matter, finding that certain effects can lead to enhanced experimental sensitivity compared to the conventional contact interaction.
For example, when the transfer momentum is sufficiently large, the time component of the dark matter current contributes significantly to the proton structure factor. 
Also, even without a light mediator, we find a residual momentum dependence in the quark-nucleon matching operation which can contribute similarly. 
We promote this endeavor by deriving direct limits on sub-GeV spin-dependent scattering of boosted dark matter from terrestrial data. 
We find that the exclusion limits from the boosted structure factor differ by as much as six orders of magnitude from those calculated using non-relativistic structure factors.

\end{abstract}
\pacs{Valid PACS appear here}
\maketitle


\section{Introduction}
Observations in cosmology and astrophysics have supported the presence of dark matter (DM)~\cite{Bertone:2004pz}. 
Its features, such as mass and interactions are however still unknown.  One of the most promising experimental 
avenues is to search for the small energy depositions from the DM elastically scattering in sensitive detectors on Earth. 
Strict constraints exist on the cross section for DM heavier than 1 GeV. 
As the detection limits reach the neutrino floor, we must look for other strategies to explore the surviving 
parameter space of DM or find some means to detect DM beyond ordinary considerations ~\cite{Battaglieri:2017aum}. 

A crucial aspect of the theoretical study of DM detection is the elastic scattering process between DM and the nuclei, 
which determines detection rates.
We can classify the hypothetical DM-nucleon interactions into the spin-independent and spin-dependent cases. 
There is already an impressive range of existing constraints on the DM-nucleon cross-section in the MeV-to-GeV mass range, 
ranging from rare processes involving the emission of photons or ‘Migdal’ electrons from the recoiling atom~\cite{Ibe:2017yqa,Dolan:2017xbu,Flambaum:2020xxo,Knapen:2020aky}, 
and the small flux of boosted dark matter (BDM) arising from interactions between DM and cosmic rays, the sun or mesons~\cite{An:2017ojc,Bringmann:2018cvk,Ema:2018bih,McKeen:2018pbb,Alvey:2019zaa,Dent:2019krz,Wang:2019jtk,Ge:2020yuf,Su:2020zny,Xia:2020wcp,Guo:2020oum,Herrera:2021puj,Bell:2021xff}. 
That said, although there is a well-developed framework for spin-independent scattering of BDM, 
a thorough treatment for spin-dependent scattering is lacking.

Since weakly interacting massive particles (WIMPs) are heavy, they are estimated to be rotating around our galaxy at a non-relativistic velocity of 
several hundred kilometers per second, and with a minimum escape velocity $\upsilon_{\rm esc}\sim 544\rm km/s$.
Therefore non-relativistic expansions of the DM-nucleon scattering amplitude in powers of 
the DM-nucleon relative velocity are appropriate for evaluating the cross-section~\cite{DelNobile:2021icc}. 
Though the non-relativistic expansion or the effective field theory is reliable in dealing with WIMP scattering, 
the neglect of higher-order terms is problematic in some circumstances. For example, 
in a novel detection of sub-GeV dark matter, called cosmic ray dark matter
(CRDM) ~\cite{Bringmann:2018cvk, Wang:2019jtk, Dent:2019krz, Feng:2021hyz,Bell:2021xff,Kolesova:2022kvq,Alvey:2022pad,Lei:2020mii}, this simple expansion is not reliable. 
This is because that the incident dark matter is boosted to be relativistic, and the momentum transfer $q$ 
is comparable to the DM mass in the sub-GeV region, thus the dropped higher-order
terms could be significant~\cite{Chang:2009yt} in the DM-nuclei scattering.

In this work, we focus on spin-dependent boosted dark matter scattering. 
We assume dark matter is a Majorana fermion for the sake of comparison. 
This is due to the fact that the Majorana-type DM always occurs in the most popular models of physics beyond the Standard Model, 
such as those featuring supersymmetry~\cite{Martin:1997ns} or extra dimensions~\cite{Csaki:2005vy}, {\em et. al.} 
It's worth noting that our calculation is not limited to Majorana DM, which can be easily extended to Dirac fermion or scalar DM. 
Previous calculations of spin-dependent WIMP scattering have started from WIMP-nucleon currents 
and used the nuclear-structure function to convert the result to the nuclei. 
However, nuclear structure calculations can be improved with recent advances in nuclear interactions and computing capabilities. 
We can thus demonstrate the difference between the WIMP-nuclei scattering and boosted DM scattering on the target.
Furthermore, we find another source of the momentum transfer effect in the contact interaction, where the finite size proton accounts for the momentum dependent behavior.  
\section{Computational Framework}
In order to demonstrate the subtlety of BDM spin-dependent scattering, we firstly sketch the ordinary computation of the WIMP spin-dependent cross section. 
At low momentum transfer $q$, the Lagrangian for the interaction between DM and quarks can be evaluated using chiral effective field theory~\cite{Klos:2013rwa,Bishara:2017nnn}. 
In the neutron or proton-only case, the differential scattering cross section can be rewritten as~\cite{XENON:2019rxp}
\begin{equation}
\frac{d\sigma^{\mathrm{SD}}}{dq^2}=\frac{\sigma_{\chi N}^{\mathrm{SD}}}{3 \mu_{N}^{2} v^2}
\frac{\pi}{2J+1} S_{A}(q),\label{eq:5_final_diff_cross_section}
\end{equation}
in which $\mu_N$ is the reduced mass of the DM-nucleon system, and $\sigma_{\chi N}$ is the scattering cross section between a DM particle and a single proton or neutron at zero momentum transfer. 
$v$ is the WIMP velocity in the rest frame of the detector, and $J$ is the initial ground-state angular momentum of the nuclei. 
The total expected non-relativistic spectrum of the detection rate  $dR/dE_{\mathrm{r}}$ is $\frac{dR}{dE_{\mathrm{r}}}= \frac{2\rho_{\chi}}{m_{\chi}} \int
\frac{d\sigma^{\mathrm{SD}}}{dq^2} v f({\bm v}) d^3 v$.  
$m_{\chi}$ is the mass of the DM, $\rho_{\chi}$ is the local DM density
and $f({\bm v})$ is the velocity distribution in the rest frame of the detector.
$q = \sqrt{2 E_{\mathrm{r}} m_{\mathrm{N}}}$ is the transfer momentum.
If we assume a  standard isothermal WIMP halo, then $v_0 = 220~\mathrm{km}/\mathrm{s}$,
$\rho_{\chi}~=~0.3~\mathrm{GeV}/(\mathrm{c}^{2}\times\mathrm{cm}^{3})$,
$v_\mathrm{esc} = 544~\mathrm{km}/\mathrm{s}$,
and Earth velocity $v_\mathrm{E} = 232~\mathrm{km}/\mathrm{s}$. 
If we generalize the WIMP into BDM, the Maxwell-Boltzmann distribution is replaced by the incoming DM flux $d\Phi_{\chi}/dE_{\chi}$. 

The structure factor $S_A(q)$ plays a crucial role in determining the event rate of spin-dependent scattering process in both relativistic and non-relativistic process,
\begin{equation}
S_A(q)=\frac{1}{4 \pi G_{\rm v}^2}\sum_ {s_f,s_i}\sum_{M_f,M_i}
\left|\left\langle {f}\left|{\mathcal L}^{\rm SD}_\chi\right|{i}\right\rangle\right|^2 \,,
\end{equation}
in which the sum $s_f, s_i = \pm 1/2$ is over Majorana fermion spin projections, and the sum $M_f, M_i$ is over the projections of the total angular momentum of the final and initial states $J_f, J_i$.  
We assume the heavy mediator is a vector so that $G_{\rm v}\sim 1/m_{\rm V}^2$.
It is easy to include the scalar mediator scenario. 
Usually, the heavy mediator does not lead to a momentum transfer effect in the scattering process. 
However, we find there is still a residual momentum transfer effect through the matching procedure from quark to proton. 
The structure factor has three contributions: the spatial current, temporal current, and interference component current. 
Fortunately, we find that the wave functions of
the spatial current and the temporal current are orthogonal to each other,
making their interference terms vanish. 
One can write the structure factor as
\begin{eqnarray}
  S_A(q)= S^0_A(q)+S_T(q),
\end{eqnarray}
where $S^0_A(q)$ denotes the contribution from the spatial current couplings and $S_T(q)$ is the time component contributions. 
Evaluating the Lagrangian density between the initial and final state, the spatial contribution is
\begin{equation}
\left\langle{f}\left| {\mathcal L}^{\rm SD}_\chi \right|{i}\right\rangle = -\frac{G_{\rm v}}{\sqrt{2}}
\int d^3{\bm r} \, e^{-i{\bm{q}} \cdot {\bm r}} \, \overline{\chi}_f
{\bm \gamma} \gamma_5 \chi_i \, {\bm J}^A_{i}({\bm r}) \,,
\end{equation}
in which  $e^{-i{\bm {q}} \cdot {\bm r}} \, \overline{\chi}_f {\bm \gamma}
\gamma^5 \chi_i =\left\langle {\chi_f}\left|{\bm j}({\bm r})\right|{\chi_i}\right\rangle $
is the matrix element of the  current of the DM and
${\bm J}^{A}_{i}({\bm r})=\sum_q A_q \overline \psi_q{\bm \gamma}\gamma_5\psi_q$ denotes the hadronic current~\cite{Klos:2013rwa}.
 For the response of the nuclei, the spin-dependent dark matter interaction couples dominantly to a single nucleon.
but also to pairs of nucleons. Then the quark currents are replaced by their expectation value in the nucleon, 
leading to 1b axial-vector current at one-nucleon level. The   DM interaction couples to nucleon pairs at order $q^3$. 
This leads to 2b axial-vector current. However,  the error of 2b current level
results is too large, Thus the 1b current results are used in this work for the simplicity.
The structure factor $S^0_A(q)$ can be decomposed as a sum over multipoles 
$L$ with reduced matrix elements of the longitudinal ${\mathcal L}^5_L$, transverse electric 
${\mathcal T}^{\mathrm{el}5}_L$, and transverse magnetic ${\mathcal T}^{\mathrm{mag}5}_L$ projections of the axial-vector currents
\begin{equation}
\begin{aligned}
S^0_A(q) &= \sum_{L \geqslant 0}\left|\langle {J_f}\!|\!|{\mathcal L}_L^5|\!|\!
{J_i}\rangle\right|^2\\
 &+\sum_{L \geqslant 1} \left(\left|\langle{ J_f}\!|\!|
{\mathcal T}_L^{\mathrm{e l}5}|\!|\!{J_i}\rangle\right|^2 
+\left|\langle{ J_f}\!|\!|{\mathcal T}_L^{\mathrm{mag}5}|\!|\!
{ J_i}\rangle\right|^2\right) \,.\label{nrs0}
\end{aligned}
\end{equation}

For fast-moving DM, the projections of the axial-vector currents will not be changed 
but become dependent on the momentum of the incident DM particle
\begin{eqnarray}
&& S^0_A(q,p_i,p_f)=\nonumber\\
&& \frac{1}{2}\biggl[\sum_{L\geqslant0}\frac{1}{E_f E_i}\left(2p_{f}^{3}p_{i}^{3}
+p_{f}\centerdot p_{i}+m_\chi^{2}\right)\left|\langle J_{f}||{\mathcal L}_L^5||J_{i}\rangle\right|^{2}\,\nonumber \\
&&+\sum_{L\geqslant1}\frac{1}{E_f E_i}\left(p_{f}^{1}p_{i}^{1}+p_{f}^{2}p_{i}^{2}+p_{f}\centerdot
p_{i}+m_\chi^{2}\right)\,\nonumber \\
&&\left(\left|\langle J_{f}||{\mathcal T}_L^{\mathrm{e l}5}||J_{i}\rangle\right|^{2}+\left|\langle J_{f}||{\mathcal T}_L^{\mathrm{mag}5}||J_{i}\rangle\right|^{2}\right)\biggl]\,.\label{eq:s0a}
\end{eqnarray}
We can also get the time component contribution to the structure factor
\footnote{ Note that we choose the time component contribution at the nucleon level to account for the
nuclear contributions to the structure factor. A detailed deviation can be found in the Appendix C.}
\begin{eqnarray}
S_{T}(q)&=&
\frac{1}{4E_{i}E_{f}}\frac{1}{4E_i^{\prime} E_f^{\prime}}\frac{1}{2\pi}\biggl[ 4\left(2E_{i}E_{f}-\frac{q^{2}}{2}-2m_{\chi}^{2}\right)\nonumber\\
&&\times\left(2E_{i}^{\prime}E_{f}^{\prime}-\frac{q^{2}}{2}-2m_{N}^{2}\right) \, \nonumber\\
&& \hspace{-7mm}+4\frac{m_{N}}{q^2+m_{\pi}^2} \left(2E_{i}E_{f}-\frac{q^{2}}{2}-2m_{\chi}^{2}\right)
\left(E_{i}^{\prime}-E_{f}^{\prime}\right) m_{N} q_{0}\nonumber\\
&& \hspace{-7mm}+\left(\frac{m_{N}}{q^2+m_{\pi}^2} \right)^{2}\left(2E_{i}E_{f}-\frac{q^{2}}{2}-2m_{\chi}^{2}\right) \frac{q^{2}}{2}
q_0^2 \biggl] \,.
\end{eqnarray}
We can make some general comments about the boosted structure factor before concluding this section. 
The spatial and time components of the structure factor for semi-relativistic kinematics 
are not a function of $q$ only, but instead the incoming dark matter momentum $p_i$ or kinetic energy $T_{\chi}$.To obtain an effective structure factor, we need to integrate out the phase space of incoming momentum
    \begin{equation}
        S_{\mathrm{eff}}(q)=\int dT_{\chi} \frac{d\Phi_{\chi}}{dT_{\chi}}S(T_{\chi},q)\,.
    \end{equation}
This is equivalent to the conventional Maxwell-Boltzmann velocity integral when the incoming DM momentum reduces to the non-relativistic regime.

\begin{figure*}[htbp]
  \centering
\subfigure[]{
\includegraphics[width=8.6cm]{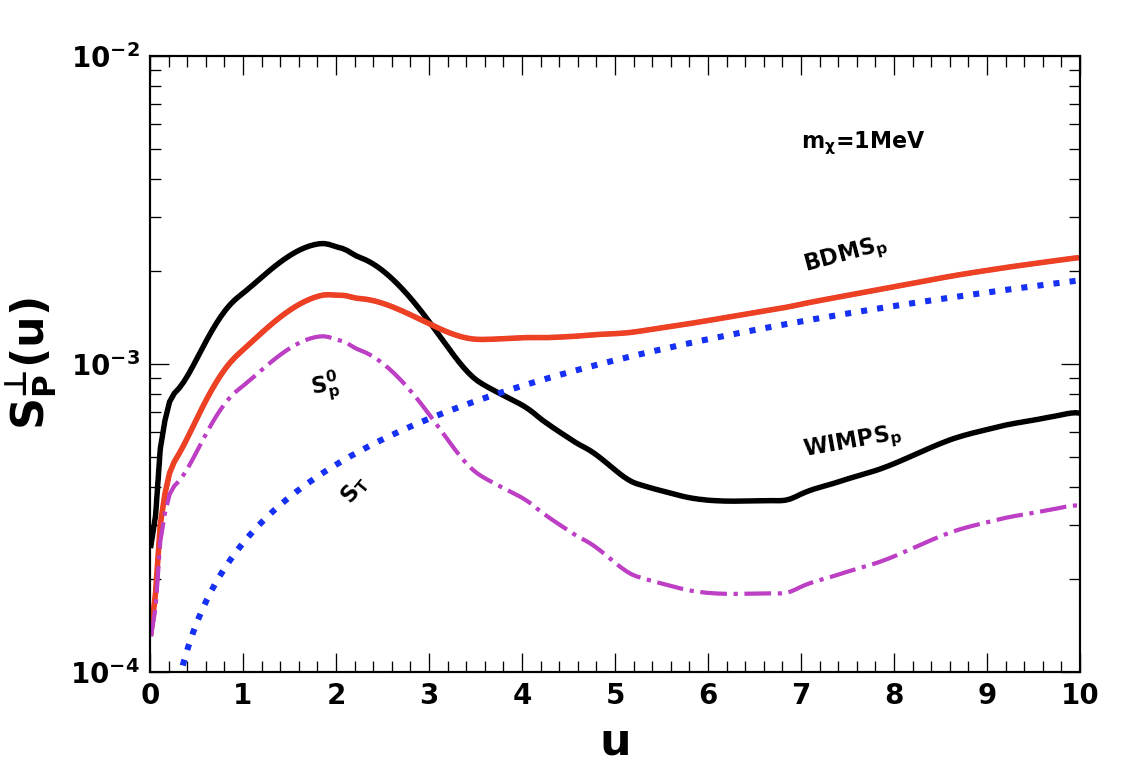}
\label{Fig1-1}
}
\subfigure[]{
\includegraphics[width=8.6cm]{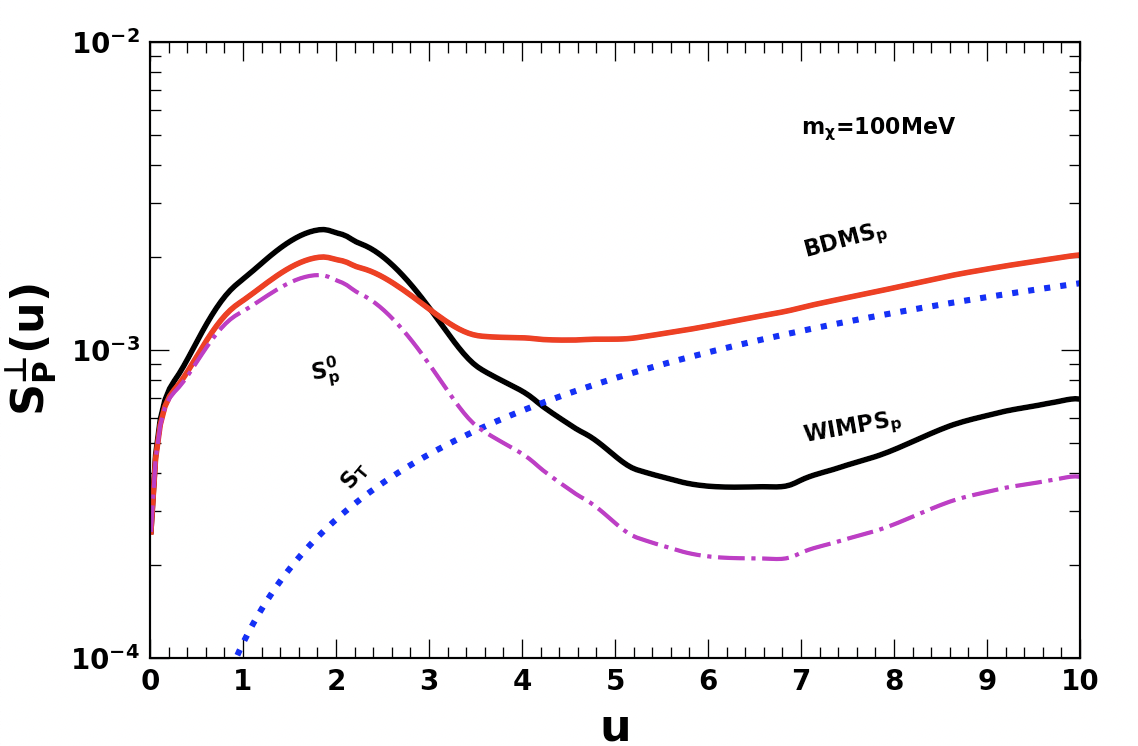}
\label{Fig1-2}
}
\subfigure[]{
\includegraphics[width=8.6cm]{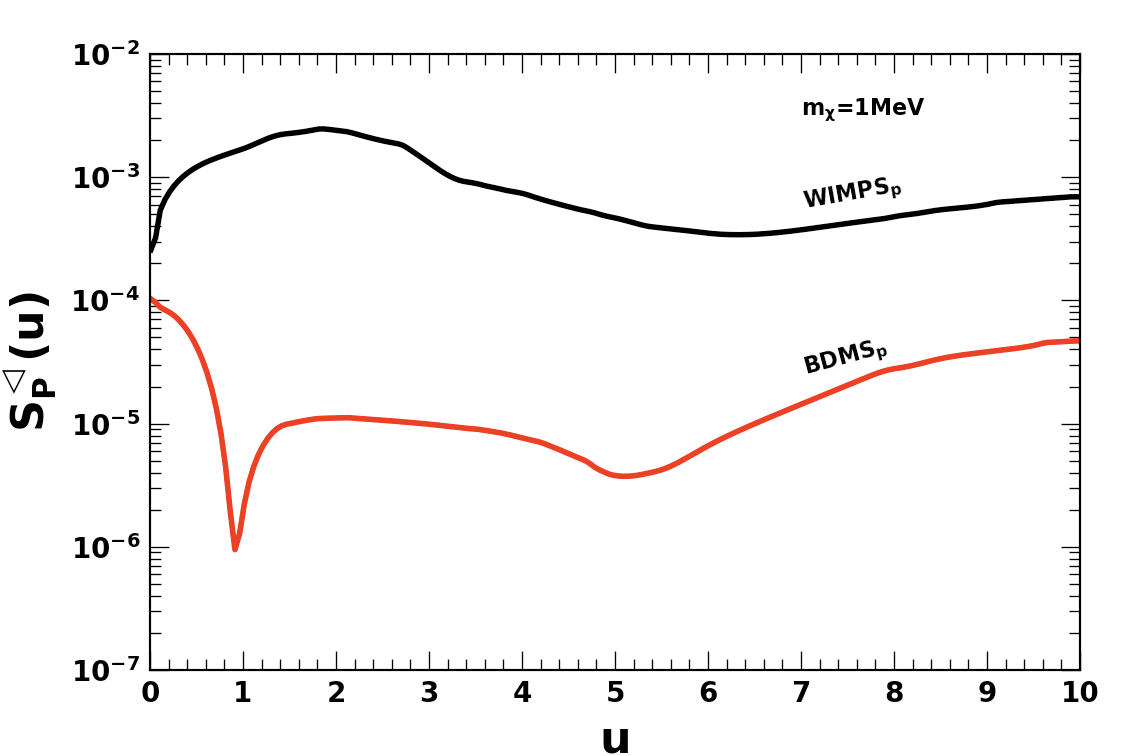}
\label{Fig1-3}
}
\subfigure[]{
\includegraphics[width=8.6cm]{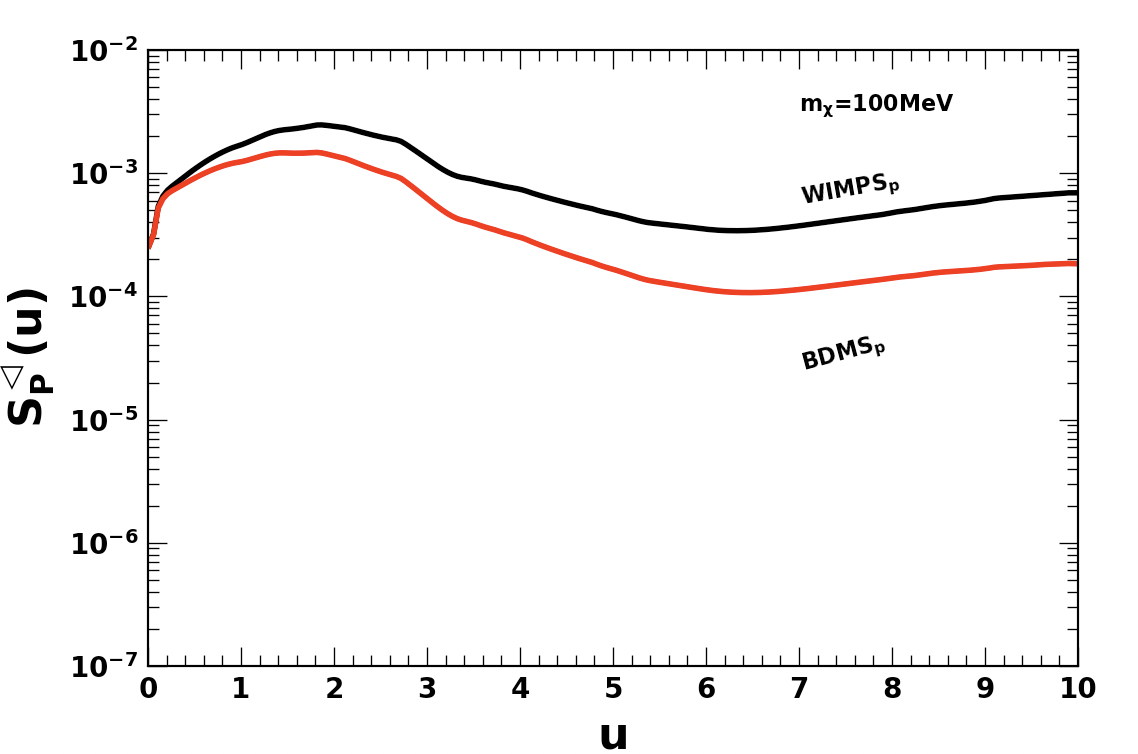}
\label{Fig1-4}
}
\caption{(Color online) Structure factors $S_{p}(u)$ for ${\rm Xe}^{131}$ as a function of $u=q^2b^2/2$, the harmonic-oscillator lengths are $b=2.2905 \rm{fm}$.
The upper two panels with $m_\chi=1 \mev$ (left ) and $m_\chi=100 \mev$ (right) 
are the results of vertical ejection of dark matter at the 1b current level. The black solid lines show the structure (WIMP) factor of the WIMP DM case.
The red solid lines show the new structure (BDM) factor obtained by considering both the time component and the spatial  contribution of the axial current in the case of boosted scattering between the DM and the nuclei. 
The pink dotted lines  are the spatial component contribution and the blue dotted lines are the time component contribution.
The lower two panels with $m_\chi=1 \mev$ (left) and $m_\chi=100 \mev$ (right) 
are the results of backward scattering at the 1b current level. The black solid lines show the structure (WIMP) factor of the WIMP DM case.
The red solid lines show the new structure (BDM) factor obtained by considering both the time component and the spatial contribution of the axial current in the case of boosted scattering between the DM and the nuclei. Note that the time component
is zero and the total contribution comes from the spatial part.}
\label{Fig:compare}
\end{figure*}

To get some feeling for this modification, we choose two specific collisions to obtain the corresponding structure factors for the 
demonstration. The first one is 
$p_{i}=(E_{i},0,0,p_{i})\,,~p_{f}=(E_{f},p_{f},0,0)\,,~
p_{i}^{\prime}=(m_{N},0,0,0)\,,~
p_{f}^{\prime}=(E^{\prime}_{f},-p_{f},0,p_{i}).$
In this case, the final direction of scattering  DM is 
perpendicular to the  the collision axis (vertical ejection). 
 And the structure factors (with a superscript tag $\perp$) are 
\begin{eqnarray}
S_A^{0\perp}(q)&=&\frac{4m_{\chi}^{2}+q^{2}}{4m_{\chi}^{2}+2q^{2}}\sum_{L\geqslant0}\left|\langle J_{f}||{\mathcal L}_L^5||J_{i}\rangle\right|^{2}\,\label{eqss0}\\
&&+\frac{4m_{\chi}^{2}+q^{2}}{4m_{\chi}^{2}+2q^{2}} 
\sum_{L \geqslant 1}\left|\langle J_{f}||({\mathcal T}_L^{\mathrm{el}5}+{\mathcal T}_L^{\mathrm{mag}5} )||J_{i}\rangle\right|^{2}\,,\nonumber
\end{eqnarray}
\begin{eqnarray}
S^\perp_{T}(q)&=&\frac{1}{8\pi}\frac{\text{1}}{2m_{\chi}^{2}+q^{2}}\frac{1}{2m_{N}^{2}+q^{2}}\,\label{eqss1}\\
&& \hspace{-5mm}\times\left[q^{4}-\frac{1}{2}\frac{q^{6}}{q^{2}+m_{\pi}^{2}}
+\left(\frac{1}{q^{2}+m_{\pi}^{2}}\right)^{2} \frac{q^{8}}{16}\right].\nonumber
\end{eqnarray}
Another collision  is 
$p_{i}=(E_{i},0,0,p_{i})\,,~p_{f}=(E_{f},0,0,-p_i)\,,~
p_{i}^{\prime}=(m_{N},0,0,0)\,,~
p_{f}^{\prime}=(E^{\prime}_{f},0,0,2p_{i}),$ 
which is the  backward scattering in case of the heavy
target nuclei. Then  the structure factors (with a superscript tag $\triangleleft$) are 
\begin{eqnarray}
S_A^{0\triangleleft}(q)&=&\frac{4m_{\chi}^2}{4m_{\chi}^2+q^{2}}\sum_{L\geqslant0}\left|\langle J_{f}||{\mathcal L}_L^5||J_{i}\rangle\right|^{2}\,\nonumber\\
&+&\sum_{L \geqslant 1}\left|\langle J_{f}||({\mathcal T}_L^{\mathrm{el}5}+{\mathcal T}_L^{\mathrm{mag}5} )||J_{i}\rangle\right|^{2}\,,
\label{eqssat}\\
S_{T}^{\triangleleft}(q)&=&0\,. \label{eqsstt}
\end{eqnarray}
The differences between non-relativistic and boosted structure factors are shown in Fig.~\ref{Fig:compare} in which 
the DM mass are chosen as $m_\chi=1 \mev$ (left ) and $m_\chi=100 \mev$
for the comparison. The upper two panels are the results of 
vertical ejection of dark matter at the 1b current level. 
The black solid lines show the structure factor of the WIMP DM.
The red solid lines show the new structure factors obtained by considering both the time component and the spatial contribution of the axial current in the case of boosted scattering 
between the DM and the nuclei. 
The pink dotted lines are the spatial component contribution and the blue dotted lines are the time component contribution.
The lower two panels are the results of backward scattering at the 1b
current level. Similar to the upper panels, the red solid lines show the new structure factor which only comes from the spatial contribution.
Form the numerical results shown in the figure, we can see that 
when the DM mass become much less than the GeV WIMP,
the structure factors are suppressed (see the left 1 MeV panels.)
in the small transfer momentum region. This is 
due to the momentum dependent coefficients in Eq.~\eqref{eq:s0a}.
Though the time component contribution
are negligible in the small transfer momentum region,
it can be dominant in the large transfer momentum
region and it can enhance the factors to a big value
when $q$ is sufficiently large.. 
Note that this time component contribution dominant region
is beyond the ordinary detection region of 
the recoil energy ($u\ll 1$) in typical DM detectors.

The momentum dependent coefficients generally suppress
the spatial contribution of the new structure factor.
The upper left panel shows that the spatial contribution 
is about one-half of the WIMP structure factor.
However, in the backward scattering, the spatial contribution of the new structure factor can be three orders of magnitude lower than WIMP structure factor. The reason can be easily derived from the momentum dependent coefficients in Eqs.\eqref{eqss0}-\eqref{eqsstt}. For 
example, the longitudinal ${\mathcal L}^5_L$ in Eq.-\eqref{eqssat}
are greatly suppressed when the DM mass is negligible compared to 
transfer momentum. This implies that the large angle scattering 
provides a much greater suppression factor than vertical eject scattering. As shown in  the following section, the  large angle scattering will become significantly important in the study of the 
boosted dark matter.

In addition, the structure factors are derived form  the four-fermion 
compact interactions. As a result, it is natural to disregard the momentum dependence of the DM i.e. $F_{\mathrm{DM}}=1$. 
In spin-dependent scattering, however, the matching between quarks and nucleons gives residual momentum dependence via pion exchange $b_1=m_N a_1/(m_{\pi}^2+q^2)$. We give a detailed calculation of the nucleon matrix elements and a detailed description of $b_1$ in Appendix C.
Such an effect does not come from the light mediator exchange, but rather the finite size effect of protons and neutrons.

\section{Benchmark Model: Cosmic-Ray Boosted Dark Matter}
In the CRDM scenario, DM is boosted by energetic galactic cosmic rays, and it subsequently becomes a fast-moving particle which is one component of  cosmic rays.
Following scattering in detectors, new limits on the DM-nucleon scattering cross section below 1 GeV can be obtained. 
CRs transfer kinetic energy to the static DM particle, making it form an energetic flux.
The DM flux in this situation resembles the neutrino flux scattering from outer space, allowing neutrino detectors, such as MiniBooNE {\it et. al. } to give constraints on the CRDM parameter space.
This relativistic DM flux can be obtained via the collision rate of CRs with DM per unit  kinetic energy of CRs ($T_i$) and DM ($T_\chi$) in a differential volume $dV$
\begin{equation}
\frac{d^{2} \Gamma_{\mathrm{CR}_{\mathrm{i}} \rightarrow \chi}}{d T_{i} d T_{\chi}}
=\frac{\rho_{\chi}}{m_{\chi}} \frac{d \sigma_{\chi i}}{d T_{\chi}}
\frac{d \Phi_{i}^{\rm{LIS}}}{d T_{i}} d V,\
\end{equation}
where the flux is taken in the local interstellar (LIS) population
of the CRs ~\cite{DellaTorre:2016jjf}, and $i$ stands for the 
specific species of the cosmic rays.
Integrating this over the relevant volume and CR energies yields a boosted DM flux
\begin{equation}
\begin{aligned}
\frac{d \Phi_{\chi}}{d T_{\chi}}&=\int_{\Omega} \frac{d \Omega}{4\pi d^2} \int_{T_{i}^{\min }} d T_{i} \frac{d^{2} \Gamma_{\mathrm{CR}_{i} \rightarrow \chi}}{d T_{i} d T_{\chi}}\\
&=D_{\mathrm{eff}} \frac{\rho_{\chi}}{m_{\chi}} \sum_{i} \int_{T_{i}^{\min }} d T_{i} \frac{d \sigma_{\chi i}}{d T_{\chi}} \frac{d \Phi_{i}^{\mathrm{LIS}}}{d T_{i}}\,.
\end{aligned}\label{eqn:flux}
\end{equation}
Since incoming proton CRs are highly relativistic, the structure factor reduces to $1$. 
When the CRDM particle travels from the upper atmosphere to the detector, the scattering with dense matter attenuates the flux to zero, which explains why CRDM searches are sometimes blind to large cross sections: large scattering cross-sections generally give a large CRDM flux, however this also leads to a significant attenuation of the flux. 
The degradation of the energy of the CRDM component can be expressed via
\begin{equation}
\frac{d T_{\chi}}{d x}=-\sum_{N} n_{N} \int_0^{E_r^{\rm max}}
 \frac{d \sigma_{\chi N}}{d E_{r}} E_{r} d E_{r}.\label{dTchidx}
\end{equation}
Here, $E_r$ refers to the energy loss by a CRDM particle in a collision with a nuclei $N$. 
${d \sigma_{\chi N}}/d E_{r}$ is the differential cross section of DM scattering on dense matter. 

Effectively, we can find the CRDM flux at the depth $z$ 
from the flux at the upper atmosphere via
\begin{equation}
{\frac{d\Phi_{\chi}}{dT_{\chi}^z} = \left(\frac{dT_{\chi}}{dT_{\chi}^z}\right)
\frac{d\Phi_{\chi}}{dT_{\chi}}
= \frac{4 m_{\chi}^2e^{z/\ell}}{\left(
2m_{\chi}+{T_{\chi}^z}  -{T_{\chi}^z} e^{z/\ell}
\right)^{2}}  \frac{d\Phi_{\chi}}{dT_{\chi}}}\,,\label{eqell}
\end{equation}
where ${d\Phi_{\chi}}/{dT_{\chi}}$ needs to be evaluated at
\begin{equation}
\label{eq:TztoT}
T_{\chi}=T_{\chi}^0(T_{\chi}^z)=
2m_{\chi} T_{\chi}^ze^{z/\ell}\left(
2m_{\chi} \!+\! {T_{\chi}^z}  \!-\! {T_{\chi}^z} e^{z/\ell}
\right)^{-1}.
\end{equation}
Here $\ell$ denotes the  mean free path of the DM particles, which can be calculated using the scattering cross section and the density of ordinary matter on Earth,
\begin{equation}
  \ell^{-1} \equiv \sum_N n_N \int_0^{E_r^{\max}} d E_r \frac{d\sigma_{\chi N}}{d E_r},
\label{freepath}
\end{equation}
Note that Eq.~\eqref{eqell} 
is valid for the attenuation with a constant cross 
section. It is only a qualitative description of the differential cross
section studied in this paper. The quantitative numerical calculation is 
implemented according to our model.

CRDM particles can transfer the energy to a target nuclei inside the detector, triggering detection events, just as with ordinary DM direct detection. 
Therefore there is a natural bridge to reinterpret existing data in the CRDM context.
The equivalence between their event rate gives rise to a constraint on the critical values of mass and coupling. 
For WIMP DM, it can read from experiment directly, while for CRDM
\begin{equation}
R =\int_{T_1}^{T_2} {d E_{\rm r}}\frac{1}{m_{\mathrm{T}}} \int_{T_{\chi}^{z,\min }}^{\infty}
d T_{\chi}^z \frac{d \Phi_{\chi}}{d T_{\chi}^z} \frac{d \sigma_{\chi T}}{d E_{\rm r}},
\label{eq:CRDMrate}
\end{equation}
where $T_1$ and $T_2$ are the analysis window for the detectors, and the DM differential flux can be regarded as a modification of the velocity distribution $f({\bm v})$. 
The differential event rate is composed of the flux and differential cross section, and the flux $d\Phi_{\chi}/dT_{\chi}^z$ is evaluated at the detector after considering attenuation processes.  
$d\sigma_{\chi T}/dE_{\rm r}$ is the differential 
cross section of DM-nuclei elastic scattering. 
Note that the mean free path $\ell$ is calculated in the integrand
for every monte carlo sample. One can easily check that the 
dominant component $S_A^0$ of the structure factor 
can be written as a function of the energy $T_\chi^z$ of the incident DM
\begin{eqnarray}
S_A^0&=& \frac{4(T_\chi^z+m_\chi)^2-q^2}{4(T_\chi^z+m_\chi)^2}
\sum_{L \geqslant 0}\left|\langle {J_f}\!|\!|{\mathcal L}_L^5|\!|\!
{J_i}\rangle\right|^2\label{eq:sacrdm}\\
 &&\hspace{-8mm}+\frac{q^2+4m_\chi^2}{4(T_\chi^z+m_\chi)^2}\sum_{L \geqslant 1} \left(\left|\langle{ J_f}\!|\!|
{\mathcal T}_L^{\mathrm{e l}5}|\!|\!{J_i}\rangle\right|^2 
+\left|\langle{ J_f}\!|\!|{\mathcal T}_L^{\mathrm{mag}5}|\!|\!
{ J_i}\rangle\right|^2\right) \,.\nonumber
\end{eqnarray}
Integrating the event rate, we can obtain a relationship between experimental data and theoretical models. 
In our numerical study, the package 
DarkSUSY~\cite{Bringmann:2018lay,Gondolo:2004sc} 
is used for simulating CRDM detection. 
The spin-dependent cross section with the new structure factor
are coded in the version DarkSUSY-6.3.1~\cite{darksusy}.
And we checked with our own modified code based on DarkSUSY-6.2.1, 
of which the results are consistent.
\begin{figure}
  \centering
  \includegraphics[width=8cm]{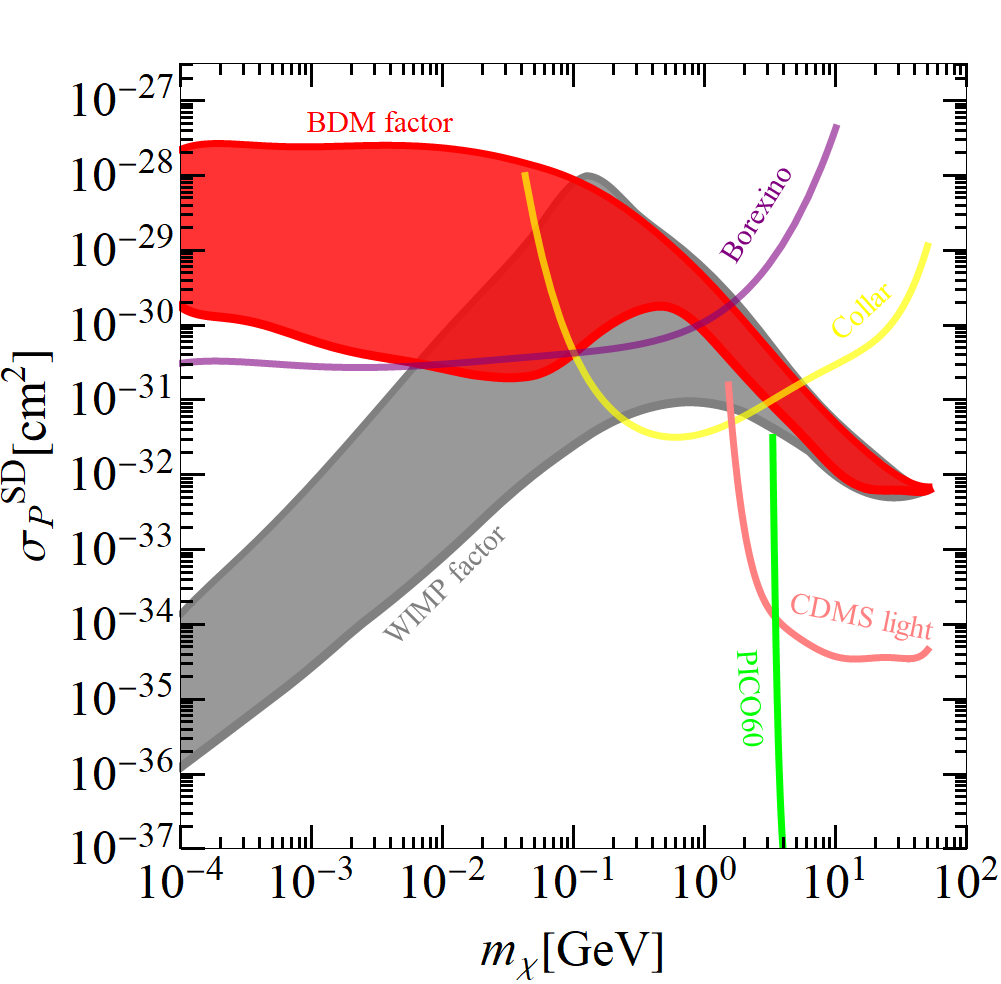}\\
 \caption{Limits on the spin-dependent cross section from Xenon1T data.
The red area shows the exclusion results using the boosted structure factor derived in this work; the dark gray contour shows the results from the non-relativistic structure factor.
The difference between the two is evidently significant, especially for light DM.
For comparison, the limits from the direct detection experiments CDMS light~\cite{SuperCDMS:2017nns}, PICO60~\cite{PICO:2017tgi} and PICASSO~\cite{IceCube:2019lus}, 
Borexino (we can refer to the purple line in FIG.3 of Ref.~\cite{Bringmann:2018cvk}) as well as from delayed-coincidence searches in near-surface detectors by Collar~\cite{Collar:2018ydf} are also shown in the plot.}.\label{SD}
\end{figure}

The numerical results of our calculation are shown in the Fig.~\ref{SD}, from which we may conclude the following. 
\begin{itemize}
\item The upper limits for the two instances considered are the same in the $m_{\chi}\gsim 0.1\rm GeV$ regime 
because the BDM structure factor recovers the WIMP structure factor when the DM mass is heavier than $1\rm GeV$. 
We can see that the exclusion regions are distinct below $0.1\rm GeV$, 
where the BDM limits become much weaker due to the suppression factor added in Eq.~\eqref{eq:sacrdm}. 
Notably, BDM limits can be as much as six orders of magnitude weaker 
than the results of WIMP structure factor below the MeV mass region. 
One can analytically derive from Eq.~\eqref{eq:s0a} that 
this much weaker limits mainly come from the large angle scattering of the CRDM, as shown in the above section.
The detailed calculations of structure factors can be found in the supplemental material.
\item It's worth noting that the detection of the recoil energy $E_{\rm r}$ necessitates a very tiny $u$, yet the transfer momentum $q$ is sufficient.
The maximum $E_{\rm r}$ (40 KeV), for example, indicates that $q$ is at about $100$ MeV. 
These results reveal that the new spatial contribution to the structure factor is critical in the evaluation of the spin-dependent scattering between DM and nuclei.
\item Another point to note is that the exclusion regions are similar whether or not there is a light mediator between DM and the target nuclei. This can be seen in the WIMP structure factor, although the boost effects on the other hand somewhat cancel it out.
The contact interaction leads to a constant cross section, and the exclusion region is horizontal (as shown 
in the results of Borexino~\cite{Bringmann:2018cvk}). 
Our results indicate two distinct regions of momentum dependence: the shape of the lower limits is comparable to those from the WIMP structure factor but with an overall difference of two orders of magnitude when $m_\chi \gsim 40\rm MeV$; below 40 MeV, the exclusion region is modified to a considerably larger value canceling the momentum dependence.

The value $m_{\chi}$ dominates the suppression factor when $m_{\chi}\gsim 40\rm MeV$, according to numerical results, whereas $T_{\chi}^z$ dominates in the $m_{\chi}\lsim 40\rm MeV$ region. 
This is owing to the fact that the typical transfer momentum $q$ is around 4$\sim$100 MeV.
This means the confined quarks in the nucleon can cause the physics of DM detection to differ significantly from the ordinary non-relativistic scenario. 
Equivalently, the finite nuclear size effects lead to momentum dependence in the scattering. 
\end{itemize}
\section{Conclusion}
The ongoing search for dark matter is of critical importance to modern physics. 
In this letter, we provide the first comprehensive treatment of spin-dependent scattering of boosted dark matter (BDM), a type of dark matter where the usual non-relativistic approximations are not reliable.

We found that when the DM is light, the spatial contribution of the boosted structure factor can be much smaller than the WIMP structure factor, while the time component contribution can boost the proton structure factor when the momentum transfer is large enough. 
We also discovered that, even in the absence of a mediator between dark matter and nuclei target, finite nuclear size effects lead to a residual momentum transfer effect. 
The complexity in the calculations of the novel boosted structure factor arises because it not only depends on the transfer of momentum $q$, but also the incoming DM momentum.

This new structure factor was applied to the CRDM scenario, providing novel 
insight into light DM detection.
In particular, we showed that the exclusion limits can differ by as much as six
orders of magnitude from those calculated using the ordinary non-relativistic
paradigm. Our findings would give conceivable hints on the future
search for the light dark matter.



\begin{acknowledgments}
We thank Nick Houston for useful discussions on the manuscript.
This work was supported by the Natural Science Foundation of China 
under grant number 12275134, 
11775012 and 11805161.
\dots.
\end{acknowledgments}

\appendix

\section{Axial-vector current of incoming dark matter particle}

We adopt the two-component spinor conventions from the  Ref.~\cite{Dreiner:2008tw} in our calculation. The general axial-vector current of Majorana fermion is given by
\begin{eqnarray}
{j}^\mu({x})&=&\overline{u}{(p_{f},s_{f})}\gamma^{\mu}\gamma^{5}u{(p_{i},s_{i})}\nonumber \\
&=&(\begin{array}{cc}
y_{f} & x_{f}^{\dagger}\end{array})\left(\begin{array}{cc}
0 & \sigma^{\mu}\\
-\overline{\sigma}^{\mu} & 0
\end{array}\right)\biggl(\begin{array}{c}
x_{i}\\
y_{i}^{\dagger}
\end{array}\biggl)\nonumber \\
&=&\left(-x_{f}^{\dagger}\overline{\sigma}^{\mu}x_{i.}+y_{f}\sigma^{\mu}
y_{i}^{\dagger}\right) \nonumber \\
&=&-\chi_{s_f}^{\dagger}\sqrt{p_f\centerdot\sigma} \overline{\sigma}^{\mu}
\sqrt{p_i\centerdot\sigma} \chi_{s_i}\nonumber \\
&&+\chi_{s_f}^{\dagger}\sqrt{p_f\centerdot
\overline{\sigma}}\sigma^{\mu}\sqrt{p_i\centerdot
\overline{\sigma}}\chi_{s_i}
\,,
\end{eqnarray}
where $x$, $x^\dagger$, $y$, and $y^\dagger$ are the two-component spinors and $s_{i,f}=\pm \frac{1}{2}$. 
The relevant basis of two-component spinors $\chi_s$ are eigenstates of $\frac{1}{2}{ \bm p}\centerdot {\bm s}$.
In the non-relativistic limit,
\begin{widetext}
\begin{eqnarray}
{j}^\mu({x})=-4s_{i}s_{f}m_{\chi}Z_{-s_{i},-s_{f}}^{\mu}(\bm{ p}_{f},\bm{p}_i)
+m_{\chi}Z^\mu_{s_{f},s_{i}}(\bm{ p}_f,\bm{ p}_i) \,,
\end{eqnarray}
with
\begin{equation}
Z_{s_i s_f}^{\mu}(\bm{p}_i,\bm{p}_f)=\Biggl\{\begin{array}{c}
\delta_{s_i s_f }+\left(\frac{\bm{p}_{i}}{2m_{i}}+\frac{\bm{p}_{f}}{2m_{f}}\right)
\,\centerdot {s}^{{a}}\tau_{s_{i},s_{f}}^{a} \qquad \qquad \qquad \qquad \qquad \mu=0\\
{s}^{a\alpha}\tau_{s_i s_f}^{a}+\left(\frac{p_{i}^{\alpha}}{2m_{i}}
+\frac{p_{f}^{\alpha}}{2m_{f}}\right)\delta_{s_is_f}+\left(\frac{p_{i}^{\beta}}{2m_{i}}
-\frac{p_{f}^{\beta}}{2m_{f}}\right)i\epsilon^{\alpha \beta \gamma }{s}^{a \gamma}\tau_{s_i s_f}^{a}
\qquad \mu=\alpha=1,2,3  \,,
\end{array}
\end{equation}
where $\tau_{}^{a}$ are the matrix elements of the Pauli matrices. We use the symbol $\tau$ rather than $\sigma$ to emphasize that the indices of the Pauli matrices $\tau^a$ are spin labels $s_i, s_f$. Then one can easily get the  time component of dark matter axial current
\begin{equation}
{j}^0({x})=-4s_{i}s_{f}m_\chi\left[\delta_{-s_{i},-s_{f}}
+\left(\frac{\bm{p}_{i}}{2m_{\chi}}+\frac{\bm{p}_{f}}{2m_{\chi}}\right)\centerdot
{s}^{a}\tau_{-s_{i},-s_{f}}^{a}\right]+m_\chi\left[\delta_{s_{i},s_{f}}
+\left(\frac{\bm{p}_{i}}{2m_{\chi}}+\frac{\bm{p}_{f}}{2m_{\chi}}\right)
\centerdot {s}^{a}\tau_{s_{i},s_{f}}^{a}\right] \,.
\end{equation}
We can see that the momentum term is subleading in case of a low velocity and the mass terms cancel each other after summation of the spins. However, it is also evident that the time component of the axial current of dark matter becomes significant in the relativistic limit
\begin{equation}
{j}^0({x})=-\chi_{s_{f}}^{\dagger}\frac{(E_f+m_{\chi})\bm{\sigma}\centerdot \bm{p_i}+(E_i+m_{\chi})\bm{\sigma}\centerdot \bm{p_f}}{\sqrt{(E_f+m_{\chi})(E_i+m_{\chi})}}\chi_{s_{i}} \,.
\end{equation}
\end{widetext}
Thus, the time component of the axial current can not be neglected in the calculation of the cross sections when the incoming dark matter is relativistic.

Next we give  the proof of orthogonality of spatial and time component. 
The time and spatial components of axial current of dark matter are shown respectively
\begin{equation}
j^0=-x_{(p_{f},s_{f})}^{\dagger}x_{(p_{i},s_{i})} \,
+y_{(p_{f},s_{f})}y_{(p_{i},s_{i})}^{\dagger} \,,
\end{equation}
\begin{equation}
j^i=x_{(p_{f},s_{f})}^{\dagger}{\bm \sigma}x_{(p_{i},s_{i})}
+y_{(p_{f},s_{f})}{\bm \sigma}y_{(p_{i},s_{i})}^{\dagger}\,.
\end{equation}
The interference of time component and spatial component
\begin{widetext}
\begin{eqnarray}
&&\frac{1}{2}\sum_{s_{i},s_{f}}(-x_{(p_{f},s_{f})}^{\dagger}x_{(p_{i},s_{i})}+y_{(p_{f},s_{f})}y_{(p_{i},s_{i})}^{\dagger})(x_{(p_{i},s_{i})}^{\dagger}{\bm \sigma}x_{(p_{f},s_{f})}+y_{(p_{i},s_{i})}{\bm \sigma}y_{(p_{f},s_{f})}^{\dagger})\nonumber\\
&=&\frac{1}{2}\sum_{s_{i},s_{f}}-x_{(p_{f},s_{f})}^{\dagger}p_{i}\centerdot\sigma{\bm \sigma}x_{(p_{f},s_{f})}-x_{(p_{f},s_{f})}^{\dagger}m_{\chi}{\bm \sigma}y_{(p_{f},s_{f})}^{\dagger}+y_{(p_{f},s_{f})}m_{\chi}{\bm \sigma}x_{(p_{f},s_{f})}+y_{(p_{f},s_{f})}p_{i}\centerdot\bar{\sigma}{\bm \sigma}y_{(p_{f},s_{f})}^{\dagger} \nonumber\\
&=&\frac{1}{2}{\rm Tr}[(-p_{f}\centerdot\sigma p_{i}\centerdot\sigma{\bm \sigma}-m_{\chi}^{2}{\bm \sigma}+m_{\chi}^{2}{ \bm \sigma}+p_{f}\centerdot\bar{\sigma}p_{i}\centerdot \bar{\sigma}{\bm \sigma})] \nonumber\\
&=&0\,.
\end{eqnarray}
\end{widetext}
\section{The spatial component of the structure factor $S^0_A(q)$}

We begin our calculation from the scattering amplitude
\begin{equation}
\langle{f}\bigl| {\mathcal L}^{\rm SD}_\chi \bigl|{i}\rangle = -\frac{G_{\rm v}}{\sqrt{2}}
\int d^3{\bm r} \, e^{-i{\bm q} \cdot {\bm r}} \, \overline{\chi}_f
{\bm \gamma} \gamma_5 \chi_i \, {\bm J}^A_{i}({\bm r}) \,.
\end{equation}
The current are expanded in terms of spherical unit vectors~\cite{Walecka:1995mi}:
\begin{equation}
\overline{\chi}_f {\bm \gamma} \gamma^5 \chi_i \, e^{-i{\bm q} \cdot {\bm r}}
= {\bm l} \, e^{-i{\bm q} \cdot {\bm r}} = \sum_{\lambda=0,\pm 1} l_\lambda
\, {\bm e}^\dagger_\lambda \, e^{-i{\bm q} \cdot {\bm r}} \,,
\label{leptonic}
\end{equation}
with spherical unit vectors with a $z$-axis in the direction of ${\bm q}$
\begin{align}
{\bm e}_{\pm1} &\equiv \mp \frac{1}{\sqrt{2}} ({\bm e}_{q1} \pm i
{\bm e}_{q2}) &{\bm e}_0 &\equiv \frac{\bm q}{|{\bm q}|} \,, \\
l_{\pm 1} &= \mp \frac{1}{\sqrt{2}} (l_1\pm il_2) &l_{\lambda=0} &\equiv l_3 \,.
\end{align}
We can also expand the product ${\bm e}^\dagger_\lambda \,
e^{-i{\bm q} \cdot {\bm r}}$ in Eq.\eqref{leptonic} in a multipole
expansion~\cite{Walecka:1995mi}. This leads to
\begin{eqnarray}
\langle{f} \bigl|\!\! \,{\mathcal L}^{\rm SD}_\chi\! \bigl|{i} \rangle
\!&=&\! -\frac{G_{\rm v}}{\sqrt{2}}\!\!\langle{J_f M_f} \bigl|\!\! \biggl( \sum_{L \geqslant 0}
\!\!\sqrt{4\pi (2L+1)}(-i)^L l_3 \, {\mathcal L}_{L0}^{5}(q)\!\nonumber \\
&&-\!\!\sum_{L\geqslant 1}\!\!\sqrt{2\pi(2L+1)}(-i)^L\!\!\sum_{\lambda=\pm 1}l_\lambda\!\!\nonumber \\
&& \!\!\left[
{\mathcal T}_{L-\lambda}^{\mathrm{e l}5}(q)+\!\lambda{\mathcal T}_{L-\lambda}^{\mathrm{mag}5}\!(q)\right]\!\!\biggl) \!\! \,\bigl|{J_i M_i}\rangle\
\label{expansion}
\end{eqnarray}
in which  $\bigl|{J_i M_i}\rangle$, $\bigl|{J_f M_f}\rangle$ denote the initial and final
states of the nuclei, $q=|{\bf q}|$.

The electric longitudinal, electric transverse, and magnetic transverse multipole operators
are defined by
\begin{eqnarray}
\hspace{-9mm}&&{\mathcal L}_{LM}^5(q) = \frac{i}{q} \int d^3{\bm r} \, \Bigl[{\bm \nabla}
\bigl[j_L(qr){\bm Y}_{LM}(\Omega_r)\bigr]\Bigr] \cdot {\bm J}^A({\bm r}) \,,
\label{multi1} \\
\hspace{-9mm}&&{\mathcal T}_{LM}^{\mathrm{el}5}(q) = \frac{1}{q} \int d^3{\bm r} \,
\bigl[{\bm \nabla} \times j_L(qr){\bm Y}^M_{LL1}(\Omega_r)\bigr]
\cdot {\bm J}^A({\bm r}) \,, \\
\hspace{-9mm}&&{\mathcal T}_{LM}^{\mathrm{mag}5}(q) = \int d^3{\bm r} \,
\bigl[j_L(qr){\bm Y}^M_{LL1}(\Omega_r)\bigr] \cdot {\bm J}^A({\bm r}) \,,
\label{multi3}
\end{eqnarray}
with spherical Bessel function $j_L(qr)$. The vector spherical
harmonics are given by
\begin{equation}
{\bm Y}^M_{LL^{\prime}1}(\Omega_r) = \sum_{m\lambda}\langle{L^{\prime}m1\lambda|L^{\prime}1LM}\rangle
Y_{L^{\prime} m}(\Omega_r) \, {\bm e}_\lambda \,.
\end{equation}
Since ${\bm J}^A({\bm r})=\sum^A_{i=1}{\bm J}^A_i({\bm r})
\delta({\bm r}-{\bm r}_i)$, the multipole operators can be written
as a sum of one-body operators:
\begin{eqnarray}
{\mathcal L}_{L M}^5(q) &=&\frac{i}{q} \sum^A_{i=1}
\Bigl[{\bm \nabla}\bigl[j_L(qr_i){\bm Y}_{L M}({\bm r}_i)\bigr]\Bigr]
\cdot {\bm J}^A_i({\bm r}_i)\\
&=&\frac{i}{\sqrt{2L+1}} \sum^A_{i=1} \Bigl[\sqrt{L+1}
j_{L+1}(qr_i){\bm Y}^M_{L(L+1)1}({\bm r}_i)\nonumber \\
&&+\sqrt{L} \, j_{L-1}(qr_i){\bm Y}^M_{L(L-1)1}({\bm r}_i)\Bigr]
\cdot {\bm J}^A_i({\bm r}_i) \,, \nonumber 
\end{eqnarray}
\begin{eqnarray}
{\mathcal T}_{L M}^{\mathrm{e l}5}(q)
&=&\frac{1}{q}\sum^A_{i=1}
\bigl[{\bm \nabla}\times j_L(qr_i){\bm Y}^M_{LL1}({\bm r}_i)\bigr]
\cdot {\bm J}^A_i({\bm r}_i)\\
&=&\frac{i}{\sqrt{2L+1}}\sum^A_{i=1}\Bigl[\sqrt{L+1} \, j_{L-1}(qr_i)
{\bm Y}^M_{L(L-1)1}({\bm r}_i)\nonumber \\
&&-\sqrt{L} \, j_{L+1}(qr_i){\bm Y}^M_{L(L+1)1}({\bm r}_i)\Bigr]
\cdot {\bm J}^A_i({\bm r}_i) \nonumber \,, 
\end{eqnarray}
\begin{eqnarray}
{\mathcal T}_{L M}^{\mathrm{mag}5}(q)
&=&\sum^A_{i=1}j_L(qr_i)
{\bm Y}^M_{LL1}({\bm r}_i)\cdot {\bm J}^A_i({\bm r}_i) \,.
\end{eqnarray}
The structure factor $S_A(q)$ is obtained from $\bigl| \langle{f}\bigl|{
\mathcal L}^{\rm SD}_\chi\bigl|{i} \rangle\bigr|^2$ by summing over the final
DM spin and over the DM final-state angular momentum
projections, and by averaging over the initial configurations.
It is thus useful to work with reduced matrix elements that do not depend on
projection numbers:
\begin{widetext}
\begin{equation}
\langle{J_f M_f}\bigl|{O}_{LM}\bigl|{J_iM_i}\rangle 
=(-1)^{J_f-M_f}\left(
\begin{array}{ccc} J_f&L&J_i \\
-M_f&M&M_i \end{array} \right)
\langle{ J_f}\!||{O}_L\!||{ J_i}\rangle \,,
\end{equation}
with $3j$ coefficients and where $O$ is a tensor operator of rank $L$.
This gives for the sum and average~\cite{Walecka:1995mi}
\begin{eqnarray}
\frac{1}{2(2J_i+1)}\!\sum_{s_f,s_i}\! \sum_{M_f,M_i}\!\!
\bigl|\!\langle{f}\bigl|\!{\mathcal L}^{\rm SD}_\chi\!\bigl|{i}\rangle\!\bigr|^2
&=& \frac{\pi G_{\rm v}^2}{(2J_i+1)} \sum_{s_f,s_i}
\biggl( \sum_{L \geqslant 0} l_3l_3^* \bigl|\langle{J_f}\!|\!|{\mathcal L}_L^5|\!|\!{J_i}\rangle\bigr|^2
\!+\!\sum_{L \geqslant 1}\biggl[\frac{1}{2}({\bf l}\cdot{\bf l}^*-l_3l_3^*)
\Bigl(\bigl|\langle{ J_f}\!|\!|{\mathcal T}_L^{\mathrm{el}5}|\!|\!{J_i}\rangle\bigr|^2 \nonumber \\
&&\hspace{-8mm}\quad+\bigl|\langle{J_f}\!|\!|{\mathcal T}_L^{\mathrm{mag}5}|\!|\!{J_i}\rangle\bigr|^2\Bigr)\!-\!\frac{i}{2}({\bf l}\times{\bf l}^*)_3
\Bigl(2 \, \text{Re}\langle{ J_f}\!|\!|{\mathcal T}_L^{\mathrm{el}5}|\!|\!{J_i}\rangle
\langle{J_f}\!||\!\!|{\mathcal T}_L^{\mathrm{mag}5}|\!|\!{J_i}\rangle^* \Bigr) \biggr]
\biggr),
\end{eqnarray}
where we have assumed that the DM spin is $1/2$, and the cross
terms vanish due to the orthogonal properties of the $3j$
coefficients.

For the sum over DM spin projections one has for $\mu,\nu=1,2,3$
\begin{eqnarray}
-\sum_{s_i,s_f}l_\mu l_\nu^*
&=& \sum_{s_i,s_f}\overline{\chi}^{s_f}(p_f)\gamma^\mu \gamma^5 \chi^{s_i}(p_i)
\, \overline{\chi}^{s_i}(p_i)\gamma^5\gamma^\nu\chi^{s_f}(p_f) \nonumber \\
&=& \sum_{s_i,s_f} \bigl(\chi^{s_f}_\delta(p_f)\overline{\chi}^{s_f}_\alpha(p_f)
(\gamma^\mu\gamma^5)_{\alpha\beta} \chi^{s_i}_\beta(p_i)
\overline{\chi}^{s_i}_\gamma(p_i)(\gamma^5\gamma^\nu)_{\gamma\delta}\bigr) ,
\end{eqnarray}
in the nonrelativistic limit
\begin{equation}
\sum_{s} \chi_\alpha^{s}(p) \overline{\chi}_\beta^{s}(p)
\approx \frac{1}{2}\left(\gamma^0+1\right)_{\alpha\beta} \,,
\end{equation}
thus
\begin{eqnarray}
-\sum_{s_i,s_f}l_\mu l_\nu^*
=\frac{1}{4}\left[2\text{Tr}(\gamma^0\gamma^\mu \gamma^5 \gamma^5 \gamma^\nu)
+2\text{Tr}(\gamma^\mu \gamma^5 \gamma^5 \gamma^\nu)\right]
=\frac{1}{2}\text{Tr}(\gamma^\mu \gamma^5 \gamma^5 \gamma^\nu) = -2\delta^{\mu\nu} \,.
\end{eqnarray}
Then
\begin{eqnarray}
\hspace{-0.2cm}\frac{1}{2(2J_i+1)}\sum_{s_f,s_i}\sum_{M_f,M_i}\bigl|
\langle{f}\bigl|{\mathcal L}^{\rm SD}_\chi\bigl|{i}\rangle\bigr|^2
&&=\frac{G_{\rm v}^2}{2}\frac{4 \pi}{(2J_i+1)}
\biggl[ \sum_{L\geqslant 0}
\bigl|\langle{J_f}\!|\!|{\mathcal L}_L^5|\!|\!{J_i}\rangle\bigr|^2\,\nonumber \\
&&+ \sum_{L \geqslant 1} \Bigl(
\bigl|\langle{ J_f}\!|\!|{\mathcal T}_L^{\mathrm{el}5}|\!|\!{J_i}\rangle\bigr|^2
+\bigl|\langle{J_f}\!|\!|{\mathcal T}_L^{\mathrm{mag}5}|\!|\!{ J_i}\rangle\bigr|^2
\Bigr) \biggr] \,.
\end{eqnarray}
Finally
\begin{equation}
S_A(q) = \sum_{L \geqslant 0} \bigl|\langle {J_f}\!|\!|{\mathcal L}_L^5|\!|\!
{J_i}\rangle\bigr|^2
 +\sum_{L \geqslant 1} \Bigl(\bigl|\langle{ J_f}\!|\!|
{\mathcal T}_L^{\mathrm{el}5}|\!|\!{J_i}\rangle\bigr|^2
+\bigl|\langle{ J_f}\!|\!|{\mathcal T}_L^{\mathrm{mag}5}|\!|\!
{ J_i}\rangle\bigr|^2\Bigr) \,.
\end{equation}
\end{widetext}
In case of a fast moving dark matter, the completeness relation is
\begin{equation}
\sum_s \chi_\alpha^{s}(p) \overline{\chi}_\beta^{s}(p) =
\Bigl(\frac{p_\mu \gamma^\mu+m_\chi}{2E_{p}}\Bigr)_{\alpha \beta} \,.
\end{equation}
Then
\begin{eqnarray}
-\sum_{s_i,s_f}l_\mu l_\nu^*
&=& \sum_{s_{i}s_{f}}\frac{p_{f\rho}\gamma^{\rho}+m_\chi}{2E_{p_{f}}}\gamma^{\mu}\gamma^{5}\frac{p_{i\sigma}\gamma^{\sigma}
+m_\chi}{2E_{p_{i}}}\gamma^{5}\gamma^{\nu}
\\
&=&\frac{1}{E_{p_{f}}E_{p_{i}}} \biggl[-p_{f}^{\mu}p_{i}^{\nu}-p_{f}^{\nu}p_{i}^{\mu}
+\left(p_{f}\centerdot p_{i}+m_\chi^{2}\right)g^{\mu\nu}\biggl]\nonumber  \,.
\end{eqnarray}
\begin{widetext}
Similarly
\begin{eqnarray}
\frac{1}{2(2J_i+1)}\sum_{s_f,s_i}\sum_{M_f,M_i}&&
\bigl|\langle{f}\bigl|{\mathcal L}^{\rm SD}_\chi\bigl|{i}\rangle\bigr|^2
=\frac{G_{\rm v}^2}{4}\frac{4\pi}{(2J_{i}+1)}\biggl[\sum_{L\geqslant0}\frac{1}{E_{p_{f}}E_{p_{i}}}\left(2p_{f}^{3}p_{i}^{3}
+p_{f}\centerdot
p_{i}+m_\chi^{2}\right)|\langle J_{f}||{\mathcal L}_L^5||J_{i}\rangle|^{2} \,\nonumber \\
&&\hspace{-0.5cm}+\sum_{L\geqslant1}\frac{1}{E_{p_{f}}E_{p_{i}}}\left(p_{f}^{1}p_{i}^{1}+p_{f}^{2}p_{i}^{2}
+p_{f}\centerdot p_{i}+m_\chi^{2}\right)|\langle J_{f}||({\mathcal T}_L^{\mathrm{e l}5}+{\mathcal T}_L^{\mathrm{mag}5})||J_{i}\rangle|^{2}\biggl] \,,
\end{eqnarray}
and
\begin{eqnarray}
S_A(q)&=&\frac{1}{4 \pi G_{\rm v}^2}\sum_ {s_f,s_i}\sum_{M_f,M_i}\bigl|\langle {f} \bigl|{\mathcal L}^{\rm SD}_\chi\bigl|{i}\rangle\bigl|^2
=\frac{1}{2}\biggl[\sum_{L\geqslant0}\frac{1}{E_{p_{f}}E_{p_{i}}}\left(2p_{f}^{3}p_{i}^{3}
+p_{f}\centerdot
p_{i}+m_\chi^{2}\right)|\langle J_{f}||{\mathcal L}_L^5||J_{i}\rangle|^{2} \,\nonumber \\
&&\hspace{-0.5cm}+\sum_{L\geqslant1}\frac{1}{E_{p_{f}}E_{p_{i}}}\left(p_{f}^{1}p_{i}^{1}+p_{f}^{2}p_{i}^{2}
+p_{f}\centerdot p_{i}+m_\chi^{2}\right)|\langle J_{f}||({\mathcal T}_L^{\mathrm{e l}5}+{\mathcal T}_L^{\mathrm{mag}5} )||J_{i}\rangle|^{2}\biggl] \,.\label{eq:apsa}
\end{eqnarray}
\end{widetext}

\begin{figure}[htbp]
  \centering
  \includegraphics[width=5cm]{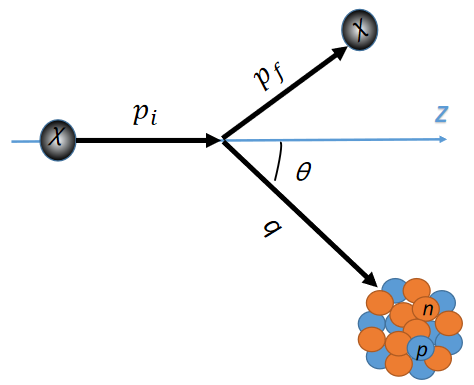}\\
 \caption{ A collision of the incident dark matter and the nuclei
 along the Z axis.}\label{xx}
\end{figure}

The collision between dark matter and nuclei 
can be chosen in a frame that the incident dark matter
with a energy $T_\chi^z$ is moving along the Z axis,
as shown in the Fig.~\ref{xx}.
The recoil direction of the nuclei can be easily got from
\begin{equation}
\cos{\theta}=\frac{q}{2p_i^3}.    
\end{equation}
Then for the isotropic incident flux, 
substituting the recoil angle $\theta$ into Eq.~\eqref{eq:apsa}
and integrating all the direction of the flux, 
the structure factor can be written as
\begin{eqnarray}
S_A(q)&=&
\frac{4(T_\chi^z +m_\chi)^2-q^2}{4(T_\chi^z +m_\chi)^2}\sum_{L\geqslant0}|\langle J_{f}||{\mathcal L}_L^5||J_{i}\rangle|^{2} \\
&&+\frac{q^2+4m_\chi^2}{4(T_\chi^z +m_\chi)^2}\sum_{L\geqslant1}|\langle J_{f}||({\mathcal T}_L^{\mathrm{e l}5}+{\mathcal T}_L^{\mathrm{mag}5})||J_{i}\rangle|^{2} \nonumber\,.
\end{eqnarray}
\section{The time component of the structure factor $S_T(q)$}\label{appen}
Time component of the Lagrangian density between the initial and final state is
\begin{equation}
\langle{f}\bigl| {\mathcal L}^{\rm SD}_\chi \bigl|{i}\rangle = \frac{G_{\rm v}}{\sqrt{2}}
 \overline{\chi}_f {\gamma}_0 \gamma_5 \chi_i  {J}^A_{0} \,.
\end{equation}
The one-nucleon time component of axial-current matrix element is~\cite{Engel:1992bf}
\begin{eqnarray}
&&\left\langle[N]p^{\prime}_f,s^{\prime}_f\left|{J}_0^A({x})\right|[N]p^{\prime}_i,s^{\prime}_i\right\rangle
=\overline{U}_{N}(p^{\prime}_f,s^{\prime}_f)\\
&&\frac{1}{2}[(a_{0}+a_{1}\tau_3)\gamma_0\gamma_{5}
+(b_{0}+b_{1}\tau_3)q_{0}\gamma_5]U_{N}(p^{\prime}_i,s^{\prime}_i)\nonumber \,,\label{rela-cur}
\end{eqnarray}
in which $U_{N}$ is a nucleon spinor, $p^{\prime}_f$, $p^{\prime}_i$ are the one-shell four momenta
and $s^{\prime}_f$, $s^{\prime}_i$ are the spin labels, $q_{\mu}=(p^{\prime}_f-p^{\prime}_i)_{\mu}$.
Note that $q^2$ is not $q_\mu q^\mu$ but $-q_\mu q^\mu$. 
The $a_0,~a_1$ are completely determined by the $A_q$ and three number $\Delta{q}$
(for $q$=$u$, $d$ and $s$ quarks) defined as
\begin{equation}
\Delta{q}s^{\mu}=\left\langle p_f,s_f\left|\overline\psi_q\gamma^{\mu}\gamma_5\psi_q\right|p_f,s_f\right\rangle \,,
\end{equation}
in which the matrix element is for the proton, and $s^\mu$ is the spin vector defined in the usual way~\cite{Jaffe:1989jz}. Specifically,
the couplings of the isoscalar part and isovector part are
\begin{eqnarray}
a_0&=& (A_u+A_d)(\Delta u + \Delta d) + 2A_s \Delta s \,, \label{a0}\\
a_1&=&(A_u-A_d)(\Delta u - \Delta d). \label{a1}
\end{eqnarray}
The b coefficients can be estimated from the partially
conserved axial-vector current (PCAC)~\cite{Chen:2021guo}, just as they are for the axial weak current. $b_0$ and $b_1$ are called isoscale and isovector coefficients, the second term is from the exchange of virtual mesons. The isoscalar mesons are heavy enough to set $b_0\simeq 0$ and pion exchange induces an isovector coefficient
\begin{equation}
b_1=\frac{m_{N}a_1}{q^2+m_{\pi}^2} \,.
\end{equation}

 Next, the time component of the axial-current matrix at nucleon level 
should be translated into the nuclear matrix elements. This simply takes the form~\cite{Walecka:1995mi}
\begin{eqnarray}
\langle{f}\bigl| {\mathcal L}^{\rm SD}_\chi \bigl|{i}\rangle = \frac{G_{\rm v}}{\sqrt{2}}
\int d^3{\bm r} \, e^{-i{\bm q} \cdot {\bm r}} \, \overline{\chi}_f { \gamma}_0 \gamma_5 \chi_i \, \rho ({\bm r}) \,,
\end{eqnarray}
in which $\rho({\bm r})$ is the charge distribution density of axial current in a nuclei. 
However, as far as we know, the axial charge distributions in the nuclei still lack experimental data.
Thus in our paper, we take the one nucleon time component contributions 
to account for the contributions at nuclear level for the simplicity. 
Summing the final state and averaging the initial state, we have
\begin{widetext}
\begin{eqnarray}
\frac{1}{4}\sum_{s^{\prime}_f,s^{\prime}_i}\sum_{s_f,s_i}\bigl|\langle{f}\bigl|{\mathcal L}^{\rm SD}_\chi\bigl|{i}\rangle\bigr|^2
&=&\frac{1}{4E_{i}E_{f}}\frac{1}{4E_i^{\prime}E_f^{\prime}}\frac{G_{\rm v}^2}{2}\biggl[(a_0+a_1 \tau_3)^2\left(2E_{i}E_{f}-\frac{q^{2}}{2}-2m_{\chi}^{2}\right)
\left(2E^{\prime}_{i}E^{\prime}_{f}-\frac{q^{2}}{2}-2m_{N}^{2}\right) \,\nonumber \\
&&+2(a_0+\tau_3)(b_0+b_1 \tau_3) \left(2E_{i}E_{f}-\frac{q^{2}}{2}-2m_{\chi}^{2}\right)
\left(E^{\prime}_{i}-E^{\prime}_{f}\right) m_{N} q_{0} \,\nonumber \\
&&+(b_0+b_1 \tau_3)^2\left(2E_{i}E_{f}-\frac{q^{2}}{2}-2m_{\chi}^{2}\right) \frac{q^{2}}{2}
q_0^2 \biggl] \,, 
\end{eqnarray}
which are referred to as ``proton-only". It is defined by the couplings $a_0=a_1=1,\tau _3=1$. Thus in the case of  protons-only,
\begin{eqnarray}
\frac{1}{4}\sum_{s^{\prime}_f,s^{\prime}_i}\sum_{s_f,s_i}\bigl|\langle{f}\bigl|{\mathcal L}^{\rm SD}_\chi\bigl|{i}\rangle\bigr|^2
&=&\frac{1}{4E_{i}E_{f}}\frac{1}{4E_i^{\prime}E_f^{\prime}}\frac{G_{\rm v}^2}{2}\biggl[4\left(2E_{i}E_{f}-\frac{q^{2}}{2}-2m_{\chi}^{2}\right)
\left(2E^{\prime}_{i}E^{\prime}_{f}-\frac{q^{2}}{2}-2m_{N}^{2}\right) \,\nonumber \\
&&+4\frac{m_{N}}{q^2+m_{\pi}^2} \left(2E_{i}E_{f}-\frac{q^{2}}{2}-2m_{\chi}^{2}\right)
\left(E^{\prime}_{i}-E^{\prime}_{f}\right) m_{N} q_{0} \,\nonumber \\
&&+\left(\frac{m_{N}}{q^2+m_{\pi}^2}\right)^2\left(2E_{i}E_{f}-\frac{q^{2}}{2}-2m_{\chi}^{2}\right) \frac{q^{2}}{2}
q_0^2 \biggl] \,, 
\end{eqnarray}
\begin{eqnarray}
S_T(q)&=\frac{1}{4 \pi G_{\rm v}^2}\sum_ {s^{\prime}_f,s^{\prime}_i}\sum_{s_f,s_i}\bigl|\langle {f} \bigl|{\mathcal L}^{\rm SD}_\chi\bigl|{i}\rangle\bigl|^2
=\frac{1}{4E_{i}E_{f}}\frac{1}{4E_i^{\prime}E_f^{\prime}}\frac{1}{2\pi}\biggl[ 4\left(2E_{i}E_{f}-\frac{q^{2}}{2}-2m_{\chi}^{2}\right)\left(2E^{\prime}_{i}E^{\prime}_{f}-\frac{q^{2}}{2}-2m_{N}^{2}\right) \,\nonumber \\
&\hspace{-7mm}+4\frac{m_{N}}{q^2+m_{\pi}^2} \left(2E_{i}E_{f}-\frac{q^{2}}{2}-2m_{\chi}^{2}\right)
\left(E^{\prime}_{i}-E^{\prime}_{f}\right) m_{N} q_{0}
+\left(\frac{m_{N}}{q^2+m_{\pi}^2} \right)^{2}\left(2E_{i}E_{f}-\frac{q^{2}}{2}-2m_{\chi}^{2}\right) \frac{q^{2}}{2}q_0\biggl]\,.
\end{eqnarray}
\end{widetext}
\bibliography{refs}

\end{document}